\documentclass[%
 reprint,
showpacs,preprintnumbers,
 amsmath,amssymb,
aps,
prl,
twocolumn
]{revtex4}

\usepackage{graphicx}% Include figure files
\usepackage{dcolumn}% Align table columns on decimal point
\usepackage{bm}% bold math
\usepackage{longtable}
\newcommand{\sg}{\ensuremath{~~\,}}

\begin{document}

%\preprint{APS/123-QED}

%% \title{Asymptotic normalization coefficients of light nuclei computed
%%   from an accurate nuclear interaction}
\title{Asymptotic normalization coefficients from {\it ab initio} calculations}
% Force line breaks with \\

\author{Kenneth M. Nollett}
 \email{nollett@anl.gov}
\author{R. B. Wiringa}%
\affiliation{%
Physics Division, Argonne National Laboratory, Argonne, IL~~60439, USA
}%

\date{\today}% It is always \today, today,

\begin{abstract}
We present calculations of asymptotic normalization coefficients
(ANCs) for one-nucleon removals from nuclear states of mass numbers $3
\leq A \leq 9$.  Our ANCs were computed from variational Monte Carlo
solutions to the many-body Schr\"odinger equation with the combined
Argonne $v_{18}$ two-nucleon and Urbana IX three-nucleon potentials.
Instead of computing explicit overlap integrals, we applied a Green's
function method that is insensitive to the difficulties of
constructing and Monte Carlo sampling the long-range tails of the
variational wave functions.  This method also allows computation of
the ANC at the physical separation energy, even when it differs from
the separation energy for the Hamiltonian.  We compare our results,
which for most nuclei are the first {\it ab initio} calculations of
ANCs, with existing experimental and theoretical results and discuss
further possible applications of the technique.

%\begin{description}
%\pacs{21.60.De} \pacs{27.10.+h}
% Ab initio methods, A ≤ 5 properties, 6 ≤ A ≤ 19 properties

%\end{description}
\end{abstract}

% Spec factors & ANCs, Ab initio methods, Quantum Monte Carlo, A ≤ 5
% properties, 6 ≤ A ≤ 19 properties

\maketitle

%\tableofcontents

Substantial experimental and theoretical effort over the past decade
and a half has been expended on the extraction of asymptotic
normalization coefficients (ANCs) from experiments involving light
nuclei
\cite{gulamov95-many,azhari01-8B,azhari01-8B-prl,beaumel01-9C,trache03-8Li,guo05-9Li,tabacaru06-8B,enders03-9C,trache02-9C}.
Most of this work has been motivated by the connection between ANCs
and astrophysical cross sections, but ANCs also offer opportunities
for significant new tests of {\it ab initio} nuclear calculations.  In
this Rapid Communication, we present predicted ANCs for several states of light
nuclei up to $A=9$, using the variational Monte Carlo (VMC) method and
a realistic Hamiltonian.

Recent years have seen
rapid advances in the {\it ab initio} theory of light nuclei
\cite{PW01,pieper08,navratil09}.  Newly-available computing power has been
brought to bear on the problem of computing properties of light ($A
\lesssim 12$) nuclei from a new generation of accurate nucleon-nucleon
and three-nucleon potentials.  
Many nuclear properties have been computed from the
modern nuclear interactions, including charge radii, electroweak
transition amplitudes, cross sections for scattering and radiative
capture, and spectroscopic factors.  Some ANCs have been computed
\cite{kievsky97,viviani05,NWS01,nollett00,navratil06}, but there has
been no systematic {\it ab initio} investigation of ANCs.

An ANC characterizes the asymptotic form of a nuclear overlap
function, which is the projection of a nuclear wave function onto a
product of subclusters.  We consider only cases of one-nucleon
removal, so the subclusters within a nucleus of mass $A$ are the
removed or ``last'' nucleon itself and a residual nucleus of mass
$A-1$.  (Although we refer to the ``last nucleon,'' our wave functions
are explicitly antisymmetric.)  The overlap channel is further
specified by orbital angular momentum $l$ and its vector sum $j$ with
the spin of the last nucleon.  The overlap function is then
\begin{equation}
\label{eqn:overlap}
R_{lj}^{J_{A-1}J_A}(r) \equiv \int \mathcal{A}\left[\Psi_{A-1}^{J_{A-1}}\left[\chi
    Y_l(\mathbf{\hat{r}})\right]_j\right]_{J_A}^\dag \frac{\delta(r-r_{cc})}{r^2}
\Psi_A^{J_A}\, 
    d\mathbf{R}\,,
\end{equation}
where $\Psi_A^{J_A}$ is the wave function of the mass-$A$ nucleus with
angular momentum $J_A$, $\Psi_{A-1}^{J_{A-1}}$ is a specific state of
the residual nucleus with angular momentum $J_{A-1}$, $\chi$ is the
spin-isospin vector of the last nucleon, and $r_{cc}$ is its
separation from the center of mass of the other $A-1$ nucleons.
Square brackets denote angular momentum coupling, $Y_l$ are spherical
harmonics, and $\mathcal{A}$ antisymmetrizes the product
$\Psi_{A-1}^{J_{A-1}}\chi Y_l$ with respect to particle exchange.  The
integral extends over all particle coordinates ${\bf R}=({\bf
  r}_1,{\bf r}_2,...,{\bf r}_A)$.

The form of the overlap as $r\rightarrow\infty$
is well known,
because it satisfies a one-body Schr\"odinger equation including at
most a Coulomb interaction.  This form contains a Whittaker function
$W_{-\eta\,m}$:
\begin{equation}
\label{eqn:anc-definition}
R_{lj}^{J_{A-1}J_A}(r\rightarrow\infty) = C_{lj}^{J_{A-1}J_A} W_{-\eta\,m}(2kr)/r\,,
\end{equation}
with $\eta = \alpha Z_{A-1}Z_N \sqrt{\mu c^2/2 B}$, $\alpha$ the
fine-structure constant, $Z_{A-1}$ and $Z_N$ respectively the charges
of the residual nucleus and the last nucleon, $\mu$ their reduced
mass, $B$ the separation energy of the last nucleon, $k=\sqrt{2\mu
  B}/\hbar$, and $m=l+1/2$.  (If the last nucleon is a neutron, then
$Z_N=0$ and $W_{-\eta\,m}(2kr)=\sqrt{2kr/\pi}K_m(kr)$, a modified
spherical Bessel function of the third kind.)  In the following, we
omit the labels $J_A$ and $J_{A-1}$ for compactness of notation.

\begin{figure}
\includegraphics[height=3.2in,angle=270]{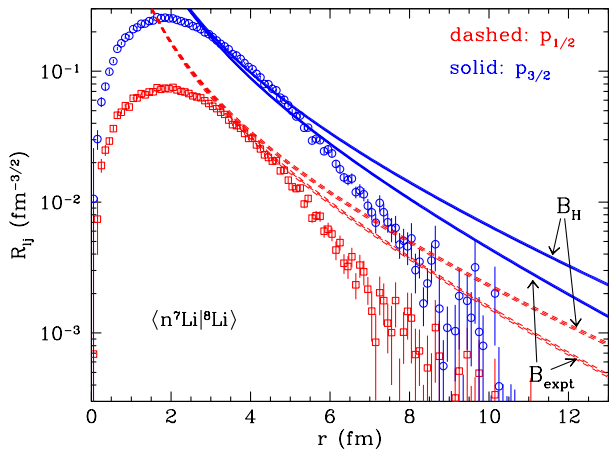}
  \caption{(Color online) Points with Monte Carlo statistical errors
    show the $^8\mathrm{Li}\rightarrow n\,^7\mathrm{Li}$ overlap,
    computed from Eq.~(\ref{eqn:overlap}), of our VMC wave functions
    in $p_{1/2}$ (red squares) and $p_{3/2}$ (blue circles)
    channels.  Curves with error bands show the asymptotic forms in
    Eq.~(\ref{eqn:anc-definition}), scaled by ANCs from
    Eq.~(\ref{eqn:anc-integral}).  Dashed (red) curves are asymptotics
    for $p_{1/2}$ and solid (blue) ones are asymptotics for $p_{3/2}$.
    They are labeled ``$B_H$'' and ``$B_\mathrm{expt}$'' according to
    the assumed neutron separation energies.}
  \label{fig:overlap}
\end{figure}

The only quantity in Eq.~(\ref{eqn:anc-definition}) that is not
determined fully by the quantum numbers and the corresponding
separation energy is the constant $C_{lj}$.  It characterizes the
overall scale of the long-range $A$-body wave function in the $lj$
channel, and it is the ANC of that channel.

It can be shown that although the spectroscopic factor $S_{lj}\equiv
\int R^2_{lj}(r)r^2 dr$ may depend strongly on the short-range
potential and the choice of wave-function representation, the ANC as
both a theoretically-computed and an experimentally-inferred quantity
is less dependent on such details \cite{friar79,akram10}.  Given
reactions (e.g., well below the Coulomb barrier) that probe only
the asymptotic part of $R_{lj}$, ANCs can be extracted from data with
fewer assumptions than spectroscopic factors can.

Computing an ANC by direct integration of Eq.~(\ref{eqn:overlap}) is
problematic for most many-body methods.  First, \textit{ab initio}
calculations may not yield the correct asymptotic form of
Eq.~(\ref{eqn:anc-definition}).  For example, methods using a
harmonic-oscillator basis have basis functions with an asymptotic form
$e^{-(r/b)^2}$, so that convergence to a long-range asymptotic form
similar to $e^{-k r}$ is slow.  In variational methods, it is often
difficult to construct a consistent set of correlations that has good
long-range asymptotics while retaining short-range properties that are
important for the variational energy.  Second, the assumed Hamiltonian
may not reproduce the experimental separation energy $B_\mathrm{expt}$
even when wave functions are computed exactly.  Third, Monte Carlo
methods suffer from the difficulty of finding a sampling scheme that
samples the tails of Eq.~(\ref{eqn:overlap}) thoroughly while
minimizing sample variance.  All three difficulties are illustrated in
Fig.~\ref{fig:overlap}.

There is another approach to computing ANCs that avoids all three of
these difficulties, and versions of it have been derived in several
contexts \cite{pinkston65,kawai67,lehman76,akram90}.  In this
approach, explicit computation of the overlap function is replaced by
an integral over the wave-function interior.
The Schr\"odinger equation
\begin{equation}
\label{eq:schroedinger}
  (H - E) \Psi_A = 0
\end{equation}
that yields wave function $\Psi_A$ with energy $E$ may be rewritten
as
\begin{eqnarray}\nonumber
\Psi_A  & = &  
  -\left[T_\mathrm{rel} + V_C + B\right]^{-1}
 \left(U_\mathrm{rel}-V_C\right)\Psi_A\\
&& - \left[T_\mathrm{rel} + V_C + B\right]^{-1}\left(H_\mathrm{int}-E_\mathrm{int}\right)\Psi_A\,.
\label{eq:greensfunction}
\end{eqnarray}
We have broken up the Hamiltonian $H$ into the relative kinetic energy
$T_\mathrm{rel}$ between the residual nucleus and last nucleon, a sum
of terms $H_\mathrm{int}$ involving only nucleons within the residual
nucleus, and a sum of terms $U_\mathrm{rel}$ involving the last
nucleon.  The point-Coulomb potential between the residual nucleus and
last nucleon is $V_C = Z_{A-1}Z_N\alpha\hbar c/r_{cc}$.  Similarly,
$E=E_\mathrm{int} -B$, with $E_\mathrm{int}$ being the purely internal
energy of the residual nucleus.

If we rewrite the Green's function $\left[T_\mathrm{rel} + V_C +
  B\right]^{-1}$ in terms of special functions, project onto the
product $\left[\Psi_{A-1}^{J_{A-1}}\left[\chi
    Y_l(\mathbf{\hat{r}})\right]_j\right]_{J_A}$ as in
Eq.~(\ref{eqn:overlap}), take advantage of the identity that
$(H_\mathrm{int}-E_\mathrm{int})\Psi_{A-1}=0$, and take the
$r\rightarrow \infty$ limit, we find that
\begin{eqnarray}
\label{eqn:anc-integral}
\lefteqn{C_{lj}=  \frac{2\mu}{k \hbar^2 w}}\\
&& \!\!\!\!\times
\mathcal{A} \int \frac{M_{-\eta\,m}(2kr_{cc})}{r_{cc}}
{ \Psi_{A-1}^\dag\chi^\dag Y^\dag_l({\mathbf{\hat{r}}_{cc}})}
\left(U_\mathrm{rel}-V_C\right) \Psi_A d\mathbf{R}\,.\nonumber
\end{eqnarray}
The integral extends over all particle coordinates, $M_{-\eta\,m}$ is
the Whittaker function that is irregular at infinity, $w$ is its
Wronskian with the regular Whittaker function $W_{-\eta\,m}$, and the 
angular momentum algebra is omitted for simplicity.  

The utility of Eq.~(\ref{eqn:anc-integral}) arises from the form of
$U_\mathrm{rel}$.  If $v_{ij}$ and $V_{ijk}$ are respectively terms of
the two- and three-body potentials involving nucleons labeled $i, j$,
and $k$, and we always label the last nucleon $A$, then
\begin{equation}
U_\mathrm{rel} = \sum_{i<A} v_{iA} + \sum_{i<j<A} V_{ijA}\,.
\end{equation}
At large separation $r_{cc}$ of the last nucleon, only the Coulomb
terms of $v_{iA}$ are nonzero.  The monopole term of their sum is
equal to $V_C$, so the factor $U_\mathrm{rel}-V_C$ in
Eq.~(\ref{eqn:anc-integral}) is short-ranged.  In our calculations, it
limits significant contributions to $r_{cc} < 7$ fm.
Equation (\ref{eqn:anc-integral}) thus reduces a problematic calculation
involving the outer regions of $\Psi_A$ to a manageable calculation
involving its interior.

We implemented Eq.~(\ref{eqn:anc-integral}) within the 
VMC method described in Ref.~\cite{wiringa09}. 
The Hamiltonian comprised the Argonne $v_{18}$ two-nucleon
\cite{WSS95} and Urbana IX three-nucleon interactions \cite{PPCW95}.
For this interaction (AV18+UIX) we constructed variational wave
functions $\Psi_A$ and $\Psi_{A-1}$ that minimize the energy
expectation values while constraining them to give approximately correct
charge radii, as determined experimentally (where known) or by exact
Green's function Monte Carlo (GFMC) calculations.  The ANC integral
was performed by Monte Carlo integration, using the same sampling
scheme (with weight proportional to $|\Psi_A|^2$) as our energy
calculations.

\begin{figure}
\includegraphics[height=3.2in,angle=270]{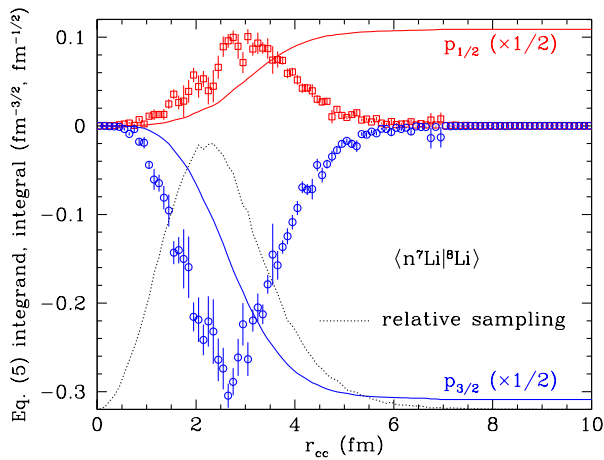}  
  \caption{(Color online)  The integrand of
    Eq.~(\ref{eqn:anc-integral}) ($\times 2\mu/k\hbar w$) is shown for
    the $p_{1/2}$ (red squares) and $p_{3/2}$ (blue circles) neutrons
    in $^8\mathrm{Li}\rightarrow n\,^7\mathrm{Li}$.  It is binned by
    the $n$-$^7$Li separation $r_{cc}$ with bars showing Monte Carlo
    errors.  The solid curves are cumulative integrals of
    Eq.~(\ref{eqn:anc-integral}), starting from the origin; at large
    $r_{cc}$, they are the ANCs (divided by 2 for visibility on this
    scale).  The dotted curve with no scale shows the distribution of
    Monte Carlo samples.}
  \label{fig:anc-integral}
\end{figure}

The distribution in $r_{cc}$ of the ANC integrand is shown in
Fig.~\ref{fig:anc-integral} for the specific case of
$^8\mathrm{Li}\rightarrow n\,^7\mathrm{Li}$.  (Where there is no
further label, the ground state of a nucleus is implied).  It may be
seen that the ANC integral is contained entirely within about 7 fm.
The distribution of Monte Carlo samples, shown as a dotted curve, is
broadly similar to the distribution of the ANC integrand, so the
integral is computed with relatively small statistical errors.

The computed $C_{lj}$ depend sensitively on the separation energies
$B$.  Equation (\ref{eqn:anc-integral}) contains $B$ implicitly through
$k=\sqrt{2\mu B}/\hbar$ and $\eta\propto 1/\sqrt{B}$, and it is
rigorously true when $B = E_\mathrm{int}-E$ for the given potential.
However, there are often significant differences between this $B$ and
the experimental separation energy $B_\mathrm{expt}$.  We computed
several ANCs in light nuclei, first using the GFMC $B_H$ for the
AV18+UIX Hamiltonian and then using $B_\mathrm{expt}$.

The use of $B_\mathrm{expt}$ in Eq.~(\ref{eqn:anc-integral}) may be
understood by considering small changes to the potential.  When $B\ll
|E|$, they can produce small changes in the wave-function interior but
large fractional changes in $B$.  The short-range part of the
variational wave function derived from AV18+UIX is, therefore, similar
to the solution that would be obtained from a slightly different
potential tuned (e.g. with small extra terms) so that
$B_H=B_\mathrm{expt}$.  Inserting a $k\propto \sqrt{B_\mathrm{expt}}$
into Eq.~(\ref{eqn:anc-integral}) matches a wave-function interior
approximating the true wave function onto the asymptotic form
corresponding to $B_\mathrm{expt}$.  Instructive illustrations of this
general principle, applied to much simpler wave functions, may be
found in Ref.~\cite{timofeyuk10}.

The use of Eq.~(\ref{eqn:anc-integral}) to compute asymptotic overlaps
is demonstrated in Fig.~\ref{fig:overlap}, where
$^8\mathrm{Li}\rightarrow\,n^7\mathrm{Li}$ overlaps computed directly
from Eq.~(\ref{eqn:overlap}) are plotted next to $C_{lj} W_{-\eta\,m}/r$
from Eq.~(\ref{eqn:anc-integral}).  It can be seen that the
$W_{-\eta\,m}$ corresponding to $B_H=1.3$ MeV \cite{pieper11} are
rather different from those for $B_\mathrm{expt}=2.03$ MeV, though both
energies are small fractions of the 41.3 MeV total binding energy for
$^8$Li.

For both $B$ values, the asymptotic $R_{lj}$ match the short-range
overlaps at $\sim 4$ fm, where the ANC integral starts to converge.
Use of $B_H$ yields $C^2_{p\, 1/2}=0.029(2)\ \mathrm{fm}^{-1}$ and
$C^2_{p\, 3/2}=0.237(9)\ \mathrm{fm}^{-1}$, compared with the
respective values $0.048(6)\ \mathrm{fm}^{-1}$ and
$0.384(38)\ \mathrm{fm}^{-1}$ from a transfer-reaction study
\cite{trache03-8Li}.  The match between the computed and ``measured''
results is poor.  Using $B_\mathrm{expt}$ yields
$0.048(3)\ \mathrm{fm}^{-1}$ and $0.382(14)\ \mathrm{fm}^{-1}$, in
very good agreement with experiment.  This pattern of agreement with
experiment for $B_\mathrm{expt}$ but disagreement for $B_H$ repeats in
all cases of substantial difference between $B_H$ and
$B_\mathrm{expt}$.  In the following, we consider only ANCs computed
from $B_\mathrm{expt}$, and we assign uncertainties based entirely on
Monte Carlo statistics rather than (difficult) assessments of the
variational wave functions.  Limited testing with variant wave
functions suggests that the total uncertainty is not much larger than
the statistical uncertainties.

\begin{figure}
\includegraphics[width=3.2in]{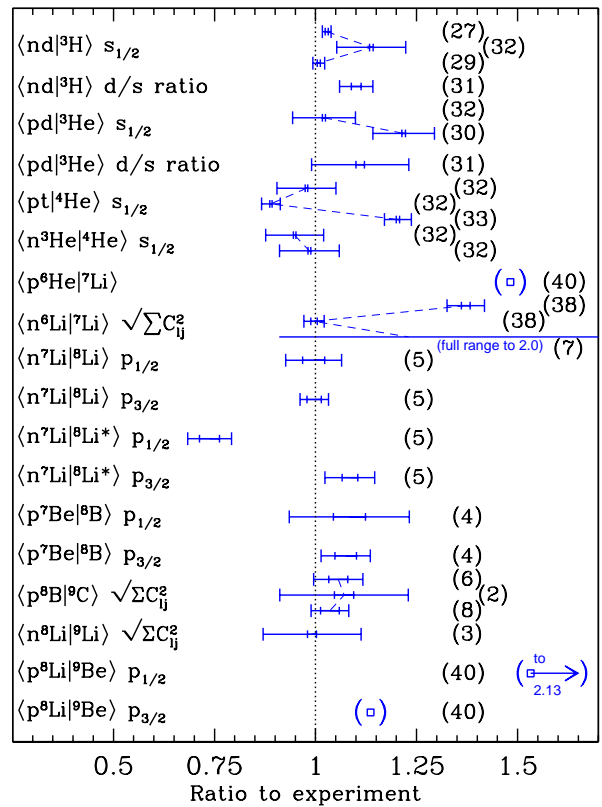}  
  \caption{(Color online) Predicted ANCs from
    Eq.~(\ref{eqn:anc-integral}), divided by experimentally-derived
    values from the references given at the right (those not appearing
    elsewhere are Refs.~\cite{purcell10,girard79,blinov85}).  For each ANC,
    small error bars indicate the Monte Carlo error of Table
    \ref{tab:combined} and larger error bars indicate its quadrature sum with
    the experimental error.  Results for the same computed ANC divided
    by different ``experimental'' numbers are joined with dashed
    lines.  Parentheses indicate particularly uncertain experimental
    constraints.}
  \label{fig:th_vs_exp}
\end{figure}

Our results are shown in Table \ref{tab:combined} and compared with
experimentally-derived numbers (where available) in
Fig.~\ref{fig:th_vs_exp}.  The lowest three sections of Table
\ref{tab:combined} repeat information from the second section, but in
``channel spin'' coupling of the form $\left[\left[J_{A-1}
    \frac{1}{2}\right]_s l\right]_{J_A}$ instead of
$\left[J_{A-1}\left[l\frac{1}{2}\right]_j\right]_{J_A}$.  We examined
most channels up to $A=9$ with either the $A$-body or the residual
nucleus in its ground state and with both stable against particle
decay.  We now comment briefly on the comparison of our results with
past work.  Extensive discussions of past experimental and theoretical
estimates may be found in Refs.~\cite{locher78,timofeyuk10}.

\begingroup
\squeezetable

\begin{table}
  \caption{ANCs computed from Eq.~(\ref{eqn:anc-integral}) for given
    $A$-body nuclei, $(A-1)$-body residual nuclei, and angular
    momentum channels $l_j$ or $^{2s+1}l$.  Units are $\mathrm{fm}^{-1/2}$, and
    $f$-wave ANCs have been multiplied by $10^3$.  Error estimates
    reflect Monte Carlo statistics only, and columns left empty are
    zero by exact symmetries. Asterisks denote first excited states.}
  \label{tab:combined}
  \begin{ruledtabular}
%    \begin{longtable}{llllll}  % use with preprint
    \begin{tabular}{llllll}   % use with twocolumn
\multicolumn{1}{c}{$A$}   & \multicolumn{1}{c}{$A-1$} & \multicolumn{1}{c}{$s_{1/2}$} & \multicolumn{1}{c}{$d_{3/2}$} &  \multicolumn{2}{c}{$C_{d\,3/2}/C_{s\,1/2}$} \\\hline
$^3\mathrm{H}$ & $^2\mathrm{H}$ & $\sg 2.127(8)$ & $-0.0979(9)$ & \multicolumn{2}{c}{$-0.0460(5)$}\\%\hline    
$^3\mathrm{He}$ & $^2\mathrm{H}$ & $\sg 2.144(8)$ & $-0.0927(10)$ & \multicolumn{2}{c}{$-0.0432(5)$}\\
$^4\mathrm{He}$ & $^3\mathrm{H}$ & $-6.55(2)$ &\\
$^4\mathrm{He}$ & $^3\mathrm{He}$ & $\sg 6.42(2)$ &\\\hline
\multicolumn{1}{c}{$A$} & \multicolumn{1}{c}{$A-1$} &\multicolumn{1}{c}{$p_{1/2}$}& \multicolumn{1}{c}{$p_{3/2}$} & \multicolumn{1}{c}{$f_{5/2}\times 10^3$} & \multicolumn{1}{c}{$f_{7/2}\times 10^3$}  \\\hline
$^7\mathrm{Li}$ &  $^6\mathrm{He}$
                  &         & $\sg 3.68(5)$ &\\
$^7\mathrm{Li}^\ast$ &  $^6\mathrm{He}$
                  &  $\sg 3.49(5)$   &         &\\
$^7\mathrm{Li}$ & $^6\mathrm{Li}$
                  & $~~1.652(12)$ & $\sg 1.890(13)$ & $-78(20)$\\
$^7\mathrm{Li}^\ast$ &  $^6\mathrm{Li}$
                  & $-0.543(16)$ & $-2.54(4)$ &\\
$^7\mathrm{Be}$ &  $^6\mathrm{Li}$
                  & $-1.87(3)$ & $-2.15(3)$ & $\sg 63(9)$\\
$^7\mathrm{Be}^\ast$ &  $^6\mathrm{Li}$
                  & $\sg 0.559(16)$ & $\sg 2.59(5)$ &\\
$^8\mathrm{Li}$ & $^7\mathrm{Li}$
                  & $\sg 0.218(6)$& $-0.618(11)$ & $~~~~5.2(5)$ & $-2.5(15)$\\
$^8\mathrm{Li}^\ast$ & $^7\mathrm{Li}$
                  & $-0.090(3)$ & $\sg 0.281(5)$ & $~\!-0.6(2)$\\
$^8\mathrm{B}$ & $^7\mathrm{Be}$
                  & $\sg 0.246(9)$ & $-0.691(17)$ & $~~~~1.1(2)$ & $-1.1(5)$\\
$^9\mathrm{C}$ &  $^8\mathrm{B}$
                  & $-0.309(7)$& $\sg 1.125(12)$ & $~~~~1.9(5)$ & $-0.5(18)$\\
$^9\mathrm{Li}$ &  $^8\mathrm{Li}$
                  & $\sg 0.308(7)$ & $-1.140(13)$ & $~\!-4.1(10)$ & $~~~\!5(3)$\\
$^9\mathrm{Li}$ &  $^8\mathrm{Li}^\ast$
                  & $-0.122(3)$  & $\sg 0.695(7)$ & $~\!-1.1(6)$\\
$^9\mathrm{Li}$ &  $^8\mathrm{He}$
                  &           & $-5.99(8)$ &           \\
$^9\mathrm{Be}$ &  $^8\mathrm{Li}$
                  & $\sg 5.03(6)$   & $\sg 9.50(11)$& $\sg 35(34)$ & $257(112)$\\
$^9\mathrm{Be}$ &  $^8\mathrm{Li}^\ast$
                  & $\sg 6.56(5)$   & $-6.21(7)$  & $~364(40)$   &\\\hline
\multicolumn{1}{c}{$A$} & \multicolumn{1}{c}{$A-1$} &\multicolumn{1}{c}{$^2p$}& \multicolumn{1}{c}{$^4p$} & \multicolumn{1}{c}{$^2f\times 10^3$} & \multicolumn{1}{c}{$^4f\times 10^3$}  \\\hline % spin channel coupling: 1/2 & 3/2
$^7\mathrm{Li}$ & $^6\mathrm{Li}$
                  & $~~2.510(18)$ & $\sg 0.029(18)$ &               & $~-78(20)$\\%x
$^7\mathrm{Li}^\ast$ &  $^6\mathrm{Li}$
                  & $-2.57(5)$ & $-0.33(3)$ &\\%x
$^7\mathrm{Be}$ &  $^6\mathrm{Li}$
                  & $-2.85(4)$ & $-0.04(4)$ &                    & $~-63(9)$\\%x
$^7\mathrm{Be}^\ast$ &  $^6\mathrm{Li}$
                  & $\sg 2.63(5)$ & $\sg 0.34(3)$ &\\%x
$^9\mathrm{Li}$ &  $^8\mathrm{Li}^\ast$
                  & $-0.599(7)$  & $ -0.373(7)$ &              & $~~~\sg 1.1(6)$\\%x
$^9\mathrm{Be}$ &  $^8\mathrm{Li}^\ast$
                  & $   -0.25(9)$   & $-9.03(8)$  &  & $-364(40)$ \\\hline
\multicolumn{1}{c}{$A$} & \multicolumn{1}{c}{$A-1$} &\multicolumn{1}{c}{$^4p$}& \multicolumn{1}{c}{$^6p$} & \multicolumn{1}{c}{$^4f\times 10^3$} & \multicolumn{1}{c}{$^6f\times 10^3$}  \\\hline  % spin channel coupling: 3/2 & 5/2
$^9\mathrm{C}$ &  $^8\mathrm{B}$
                 & $\sg 0.868(14)$& $\sg 0.779(12)$ & $~~~~0.1(19)$ & $~~\!-2(1)$\\%x
$^9\mathrm{Li}$ &  $^8\mathrm{Li}$
                 & $-0.882(15)$ & $-0.785(12)$ & $~~~~3.3(34)$ & $~~~~~5.2(19)$\\%x
$^9\mathrm{Be}$ &  $^8\mathrm{Li}$
                  & $~10.75(12)$   & $-0.25(10)$& $~256(117)$ & $~\sg 42(65)$\\%x
\hline
\multicolumn{1}{c}{$A$} & \multicolumn{1}{c}{$A-1$} &\multicolumn{1}{c}{$^3p$}& \multicolumn{1}{c}{$^5p$} & \multicolumn{1}{c}{$^3f\times 10^3$} & \multicolumn{1}{c}{$^5f\times 10^3$}  \\\hline  % spin channel coupling: 1 & 2
$^8\mathrm{Li}$ & $^7\mathrm{Li}$
                  & $-0.283(12)$& $-0.591(12)$ & $~\!-0.3(16)$ & $~~\!-5.8(10)$\\%x
$^8\mathrm{Li}^\ast$ & $^7\mathrm{Li}$
                  & $\sg 0.220(6)$ & $\sg 0.197(5)$ & & $~~~\sg 0.6(2)$\\%x
$^8\mathrm{B}$ & $^7\mathrm{Be}$
                  & $-0.315(19)$ & $-0.662(19)$ & $~\! -0.6(5)$ & $~~\!-1.4(4)$\\
%    \end{longtable}
    \end{tabular} % use with twocolumn
  \end{ruledtabular}
\end{table}
\endgroup

The $s$-wave ANCs for $A\leq 4$ nuclei have typically been inferred
from cross sections using techniques based on analyticity of the
scattering amplitude \cite{locher78,blokhintsev77}, mostly thirty or
more years ago.  Although our ANCs agree with many of those results,
Fig.~\ref{fig:th_vs_exp} demonstrates the considerable systematic
uncertainties of those methods discussed in
Refs.~\cite{friar88,locher78}.

ANCs of $^3$H and $^3$He have been computed previously from modern realistic
interactions using Eq.~(\ref{eqn:anc-integral}) 
\cite{kievsky97} and were the focus of much activity following the
development of Faddeev methods \cite{friar88,weller88,george93}.
Particular emphasis was placed on the ratio $C_{d\,3/2}/C_{s\,1/2}$,
most precisely inferred from tensor analyzing powers \cite{george93};
those results are in reasonable agreement with ours.

The Pisa group has computed ANCs for $A\leq 4$
\cite{kievsky97,viviani05} with AV18+UIX.  Their $A=3$ $C_{s\,1/2}$
are within 0.5\% of ours, but their $C_{d\,3/2}/C_{s\,1/2}$ are 10\%
smaller.  Their $^4$He ANCs are also about 6\% smaller than ours.
The reason for this difference is unclear; it
%appears too large to arise entirely from
%convergence problems in their wave functions.  Some of the difference
could reflect shortcomings of the variational wave functions, which
miss the true AV18+UIX binding energy by 850 keV in $^4$He.  Ongoing
work to compute overlaps using essentially exact wave functions from
the GFMC method seems to support our values of the $A=4$ ANCs
\cite{brida11}.  (Nuclei with $A=3,4$ have substantially identical
ANCs for $B_H$ and $B_\mathrm{expt}$ because the AV18+UIX interaction
was tuned to have $B_H \simeq B_\mathrm{expt}$ in these systems.  Pisa
ANCs converted to our conventions may be found in
Ref.~\cite{timofeyuk10}.)

For $A>4$ ANCs, experimental constraints have been inferred almost
entirely from transfer
\cite{gulamov95-many,azhari01-8B,azhari01-8B-prl,beaumel01-9C,trache03-8Li,guo05-9Li,tabacaru06-8B,goncharov87},
knockout \cite{enders03-9C}, or breakup \cite{trache02-9C} reactions,
and are of generally more recent vintage than the $A\leq 4$ ANCs.  In
some cases components of different $j$ contribute indistinguishably to
differential cross sections, which then constrain only the sum
$\sum_jC_{lj}^2$.  These cases are indicated in
Fig.~\ref{fig:th_vs_exp} and shown as the square root of the sum for
comparability of error bars.  Our $p$-shell ANCs are in broadly good
agreement with those inferred from experiment, particularly for the
well-measured $A=8$ ground state ANCs as discussed above.  (Our
calculations for $A=8$ also agree with prior theoretical estimates of
\cite{halderson04,navratil06}.)  Reference \cite{timofeyuk10} presented
many ANCs computed by applying Eq. (\ref{eqn:anc-integral}) with a
simpler potential to harmonic-oscillator wave functions derived from
shell models; about half of our $p$-shell ANCs disagree with those
calculations by more than 25\%.

The most significant differences from previous work are in the
$^7\mathrm{Li}\rightarrow n\,^6\mathrm{Li}$ ANCs.  The comparison with
experiment here is difficult because of the wide range of estimates,
which extend from $\sqrt{\sum C^2_{lj}}=1.26$ to
$2.82\ \mathrm{fm}^{-1/2}$ just from $(d,t)$ at varying energy
(\cite{gulamov95-many}, with full range shown in
Fig. \ref{fig:th_vs_exp}) and include other values within that range
\cite{goncharov87,bekbaev91}.  The effective ANC of Huang et
al. \cite{huang10}, whose capture model successfully matches
$^6\mathrm{Li}(p,\gamma)^7\mathrm{Be}$ data, is 25\% below ours.

The theoretical ANCs for $^7\mathrm{Li}\rightarrow n\,^6\mathrm{Li}$
(from a simpler model) in Ref. \cite{timofeyuk10} are 20\% to 40\%
smaller than ours.  As with $^4$He, ongoing GFMC work (with an
improved three-body interaction) seems to support our results
\cite{brida11}.  We also disagree with earlier integral-method
predictions of the ratio of $^7\mathrm{Be}\rightarrow
p\,^6\mathrm{Li}$ to isospin-mirror $^7\mathrm{Li}\rightarrow
n\,^6\mathrm{Li}$ ANCs \cite{timofeyuk03}, finding 1.15 instead of
1.05 (though we agree with their 1.12 as the ratio of $^8$B to $^8$Li
ANCs).  The sources of these differences are unclear.

Table \ref{tab:combined} includes ANCs for both $p$- and $f$-wave
channels of $p$-shell nuclei.  The small $f$-wave components arise
from the tensor terms the Hamiltonian, analogously to the $d$-wave
components in $s$-shell nuclei \cite{eiro90}.  We are unaware of any
previous calculations of $f$-wave ANCs or attempts to measure them.  A
DWBA calculation of tensor analyzing powers in sub-Coulomb
$^{208,209}\mathrm{Pb}(^7\mathrm{Li},^6\mathrm{Li})X$ (analogous to
triton $d/s$ ratio experiments) suggests that both cross sections and
analyzing powers may be too small to allow measurement of the $f/p$
ratio \cite{pieper11}.  Nonetheless, the $f$-wave ANCs demonstrate
something of the power of the integral method: Within the VMC
approach, computing ANCs for these small-amplitude channels from
Eq.~(\ref{eqn:overlap}) would require far more computing time to
achieve the same statistical accuracy, even if our variational wave
functions guaranteed the correct asymptotic form.

Several extensions of this technique within the context of quantum
Monte Carlo methods suggest themselves.  The overlaps need not
correspond only to one-nucleon removal, but may include cluster
overlaps like $^4\mathrm{He}\rightarrow dd$ and
$^7\mathrm{Be}\rightarrow \alpha\,^3\mathrm{He}$.  A straightforward
extension of the definition of ANCs to include unbound states allows
the prediction of energy widths from the integral method
\cite{kadmenskii73,akram99,esbensen01}.  The ANC integral can also be evaluated
within the GFMC method, which provides essentially exact results for a
given potential.  Use of the (computationally more demanding) Illinois
three-body potentials \cite{PPWC01} to generate the wave function
and/or the ANC kernel will provide more accurate ANCs and
$B_H$ closer to $B_\mathrm{expt}$.  Finally, use of
Eq.~(\ref{eq:greensfunction}) away from the $r\rightarrow\infty$ limit
should allow more accurate calculations of overlaps at all radii
\cite{pinkston65,kawai67,akram90b,timofeyuk10}.

\acknowledgments

We acknowledge useful discussions with I. Brida, S. C. Pieper,
A. M. Mukhamedzhanov, H. Esbensen, and C. R. Brune.  This work was
supported by the U.S.  Department of Energy, Office of Nuclear
Physics, under contract No. DE-AC02-06CH11357.  Calculations were
performed on the Fusion computing cluster operated by the Laboratory
Computing Resource Center at Argonne.

\end{document}